\documentclass{svproc}
\usepackage{url}

\usepackage{graphicx}
\usepackage{multicol}
\usepackage{footmisc}
\usepackage{amsmath,amssymb,amsfonts}
\usepackage{lineno}
\usepackage{xcolor}

\begin{document}
\mainmatter              % start of a contribution
\title{Exploring Long-Range Interactions in the Atmospheric Neutrino Oscillations at IceCube DeepCore}
\titlerunning{Long-Range Interactions}  
\author{Gopal Garg\footnote[2]{also at Institute of Physics, Sachivalaya Marg, Sainik School Post, Bhubaneswar 751005, India and Department of Physics, Aligarh Muslim University, Aligarh 202002, India.} \\
(For the IceCube Collaboration\footnote[1]{\protect\url{http://icecube.wisc.edu}})}
\authorrunning{Gopal Garg (For the IceCube Collaboration)} % abbreviated author list (for running head)
\tocauthor{Gopal Garg}
\institute{Dept. of Physics and Wisconsin IceCube Particle Astrophysics Center,\\
University of Wisconsin-Madison, Madison, WI 53706, USA\\
\email{ggarg@icecube.wisc.edu}}

\maketitle              % typeset the title of the contribution

\begin{abstract}
The IceCube neutrino observatory consists of an array of Digital Optical Modules (DOMs) instrumenting one cubic-kilometer of deep glacial ice at the South Pole. DeepCore, a densely-spaced sub-array of DOMs at the bottom central region of IceCube, enables the detection of atmospheric neutrinos with an energy threshold in the GeV range. The high statistics data of DeepCore provides a unique opportunity to perform neutrino oscillation studies as well as explore various sub-leading Beyond the Standard Model (BSM) physics signatures. We consider a well-motivated minimal extension of the Standard Model by an additional anomaly-free, gauged lepton-number symmetry, such as $L_e - L_\mu$  or $L_e - L_\tau$. These symmetries give rise to flavor-dependent long-range interaction mediated through a very light neutral gauge boson. In this contribution, we present the sensitivity of the IceCube DeepCore detector to search for this flavor-dependent long-range interaction potential with a runtime of 9.3 years.

\keywords{BSM Physics, Neutrino oscillations, IceCube DeepCore}
\end{abstract}
\section{Introduction}
\begin{figure}
    \centering
    \includegraphics[width=\linewidth]{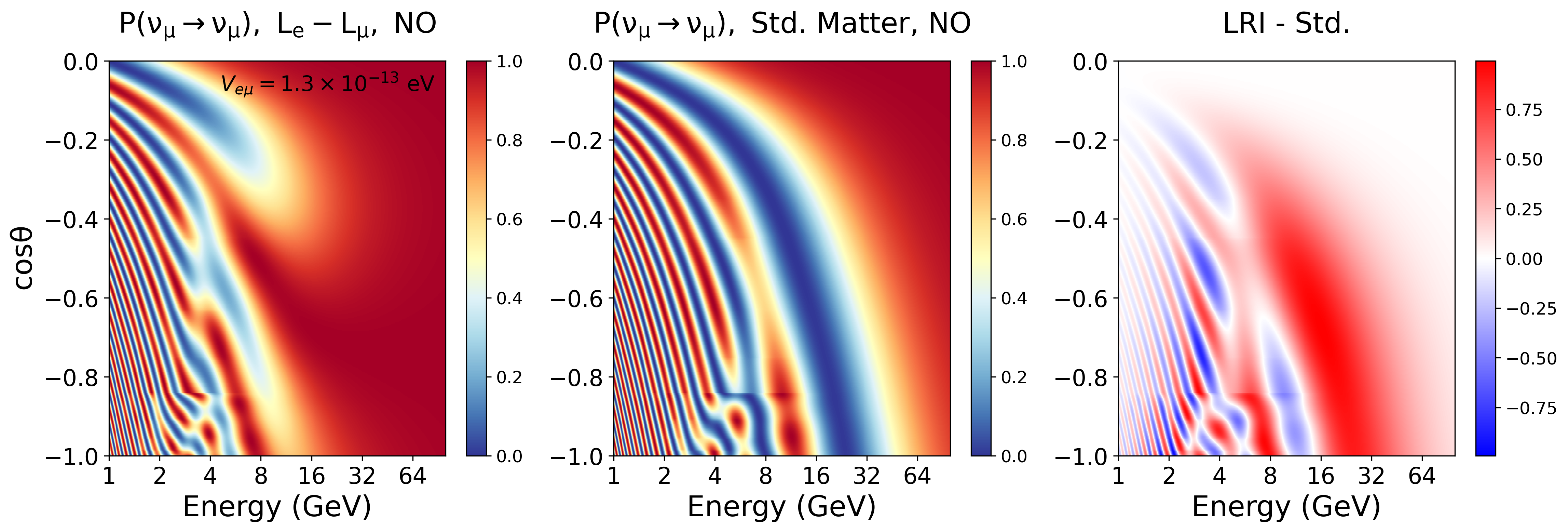}
    \caption{Three-flavor $P(\nu_\mu \rightarrow \nu_\mu)$ oscillogram in Energy and $\rm cos \theta$ plane. The left, middle, and right panels correspond to LRI potential $V_{e\mu} = 1 \times 10^{-13}$ eV, standard case, and the difference between these two scenarios, respectively.}
    \label{fig:emu_oscillogram}
\end{figure}

The neutrino flavor oscillation phenomenon compels us to extend the Standard Model (SM) to accommodate neutrino mass and mixing. Here, we consider a well-motivated minimal anomaly-free Standard Model extension through two lepton number symmetry, $L_e - L_\mu$ or $L_e - L_\tau $, sourced by the electrons~\cite{SK:2004,KamLand:2007}. We can gauge only one symmetry at a time, which introduces a new neutral $Z'_{e\mu}$ or $Z'_{e\tau}$ gauge boson, giving rise to an additional flavor-dependent, vector-like, leptonic long-range interaction. This new interaction modifies neutrino oscillations and can be tested in experiments like DeepCore.\\
In principle, this extra gauge boson can be very heavy or ultra light, which couples very feebly with the matter~\cite{PRL:2018}. Here, we consider this gauge boson to be very light, having a mass corresponding to the range equal to Sun-Earth distance or more so that it can source the Long-Range Interaction (LRI) potential from the huge electronic contents present inside the Sun and Galaxies~\cite{Pragyan:2024,Masoom:2023,Sudipta:2023,Amina:2018}. This LRI potential depends on the electron number density inside the range of the interaction and the mass of the mediating gauge boson.\\
For example, if we consider the range of this interaction to be Sun-Earth distance $(R_{\rm SE} \approx 1.5 \times 10^{13} ~\rm cm = 7.6 \times 10^{26} ~\rm GeV^{-1})$ then the $N_e ~(\approx 10^{57})$ numbers of Solar electrons can generate a long-range potential $V_{e\mu / e\tau}$ at the Earth's surface as,
\begin{equation}
 V_{e\mu/e\tau} (R_{SE}) = \alpha_{e\mu/e\tau} \frac{N_e}{R_{SE}}
	\approx 1.3 \times 10^{-11}  \rm eV (\frac{\alpha_{e\mu/e\tau}}{10^{-50}}),
\end{equation}
where $\alpha_{e\mu/e\tau} = \frac{g^2_{e\mu/e\tau}}{4 \pi}$ and $g_{e\mu/e\tau}$ is the coupling strength of the symmetry.
In the presence of LRI, the neutrino Hamiltonian in the flavor basis gets modified as, 
\begin{equation}
H_f=\left(U\left[
\begin{tabular}{c c c} 0 & 0 & 0\\
                      0 & $\frac{\Delta m^2_{21}}{2E}$ & 0\\
                      0 & 0 & $\frac{\Delta m^2_{31}}{2E}$
\end{tabular}\right]U^\dag\, + \left[
\begin{tabular}{c c c} $V_{CC}$ & 0 & 0\\ 0 & 0 & 0 \\ 0 & 0 & 0
\end{tabular} \right] + \left[
\begin{tabular}{c c c} $\zeta$ & 0 & 0 \\ 0 & $\xi$ & 0\\ 
                        0 & 0 & $\eta$ \\
\end{tabular} \right]\right),\
%\label{eq:Hf}
\end{equation}
where the first two terms are conventional vacuum and standard matter potential terms with the PMNS mixing matrix U and mass-splitting $\Delta m_{ij}^2 = m^2_i - m^2_j$, while the third term is due to the LRI. For $L_e - L_\mu$ symmetry, $\zeta = V_{e\mu}$, $\xi = -V_{e\mu}$, and $\eta = 0$ while for $L_e - L_\tau$ symmetry, $\zeta = V_{e\tau}$, $\eta = -V_{e\tau}$, and $\xi = 0$.\\
In Fig. \ref{fig:emu_oscillogram}, we show the effect of $L_e - L_\mu$ symmetry on the $\nu_\mu \rightarrow \nu_\mu$ oscillogram for illustration purposes. The left panel shows the $\nu_\mu \rightarrow \nu_\mu$ oscillogram in the presence of LRI potential with $V_{e\mu} = 1 \times 10^{-13}$ eV, the middle panel shows the oscillogram for the standard case, and the right panel shows the difference between these two scenarios. We can notice that the LRI sensitivity mainly comes from higher baselines and intermediate energy regions.

In this study, we present the capability of DeepCore to constrain LRI potential introduced by $L_e - L_\mu$ or $L_e - L_\tau$ symmetry using 9.28 years of simulated atmospheric neutrino events. Later, we will perform this study using a real dataset.

\section{IceCube DeepCore Detector}
IceCube is a 1 $\rm km^3$ neutrino observatory located at the South Pole~\cite{IceCube:2011}, which uses 5160 digital optical modules (DOMs) deployed like beads on 86 vertical cables at a depth of approximately 1.5 km to 2.5 km inside the Antarctic ice. These DOMs use photomultiplier tubes to detect the Cherenkov radiation emitted by secondary charged particles produced in the interaction of neutrinos within the ice, which is then converted into an electronic signal with an associated electronic circuit. DeepCore is the bottom central region of IceCube, which has closely spaced DOMs and seven additional strings. The denser geometry of DeepCore enables the detection of atmospheric neutrinos with energies of a few GeV, which can provide precise measurements of the mixing angle $\theta_{23}$, mass splitting $\Delta m_{32}^2$, sign of $\Delta m_{32}^2$, and constraints on BSM models like neutrino decay, NSI, LRI, etc.

\section{Analysis Details}
In this sensitivity study, we use 9.28 years of Monte Carlo event sample having around 165k events binned in twelve logarithmic bins for the reconstructed energy ($\in$[5, 100] GeV) and ten linear bins for reconstructed cosine zenith ($\in$[-1, 0]). A few additional cuts are then applied to reduce the atmospheric muon background and detector noise, resulting in a neutrino-dominated sample, followed by CNN-based reconstruction. These MC events are further classified into three Particle Identification (PID) bins based on their signatures in the detector as cascade $(0 < \rm PID < 0.25)$, mixed $(0.25 < \rm PID < 0.55)$, and tracks $(0.55 < \rm PID < 1.0)$. These MC events have been weighted by the atmospheric flux, cross sections, oscillation probabilities, and detector-related effects. We also treat the systematic uncertainties in flux, oscillation parameters, cross-section, and detector properties as nuisance parameters~\cite{IceCube:2024,IceCube:2023}. We estimate the LRI ($V_{e\mu} ~\rm or ~V_{e\tau}$) sensitivity of DeepCore using Gaussian $\chi^2$ as defined in Ref.~\cite{IceCube:2023}.

\begin{figure}
    \centering
    \includegraphics[width=0.49\linewidth]{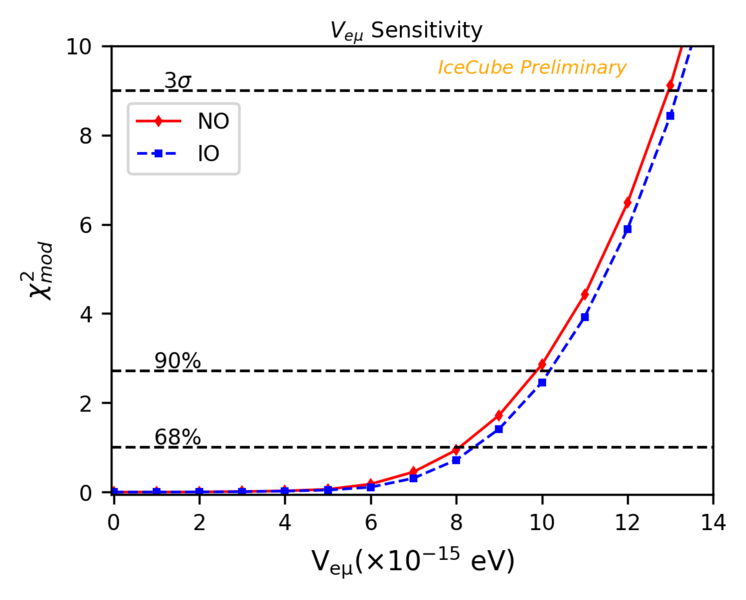}
    \includegraphics[width=0.49\linewidth]{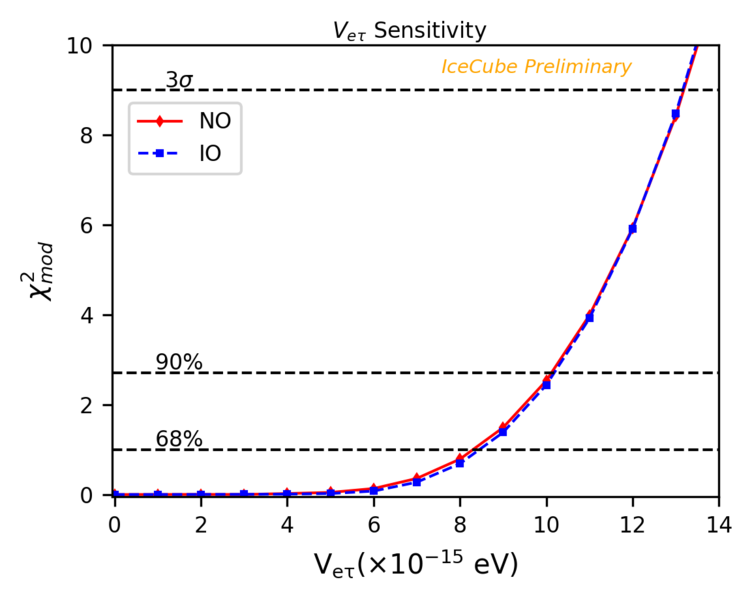}
    \caption{The preliminary sensitivity of DeepCore for $V_{e\mu}$ ($V_{e\tau}$) on the left (right) panel. The red (blue) curve corresponds to normal (inverted) mass ordering.}
	\label{fig:results}
\end{figure} 

\section{Sensitivity Results}
The left (right) panel of Fig.~\ref{fig:results} represents the Asimov sensitivity of DeepCore for the LRI potential corresponding to $L_e - L_\mu$ ($L_e - L_\tau$) symmetry using the simulated data equivalent to 9.28 years of observations. We present our sensitivity results by generating the simulated data without LRI and fitting it with a nonzero value of $V_{e\mu/e\tau}$ using the $\chi^2$ method. We obtain the 90\% upper bound on LRI potental $V_{e\mu}$ ($V_{e\tau}$) to be $9.82 \times 10^{-15}$ ($10.12 \times 10^{-15}$) eV.

\section{Conclusion}
In this study, we have presented that IceCube DeepCore can put constraints on the flavor-dependent long-range interaction, sourced by gauging $L_e - L_\mu$ or $L_e - L_\tau$ symmetry using 9.28 years of simulated data. 
 
\subsubsection{\ackname}  We acknowledge the financial support from the Department of Atomic Energy (DAE), Govt. of India, and the INSPIRE Fellowship provided by the Department of Science and Technology (DST), Govt. of India. 

%
%===============================
\bibliographystyle{spphys}

%===============================
\end{document}